\documentclass[%
 reprint,
 superscriptaddress,
 nofootinbib,
 amsmath,amssymb,
 aps,
 prx,
]{revtex4-2}

\usepackage[hidelinks]{hyperref}
\usepackage{graphicx}
\usepackage{dcolumn}
\usepackage{bm}
\usepackage{booktabs}
\usepackage{amsmath,amssymb,amsbsy,amstext, amsthm, amsfonts}
\usepackage{color}
\usepackage{slashed}
\usepackage[dvipsnames]{xcolor}
\usepackage{dsfont}
\usepackage{verbatim}
\usepackage[utf8]{inputenc} 
\usepackage{ragged2e}
\usepackage{mathtools}
\usepackage{xfrac}
\usepackage{blkarray}

\def\Vhrulefill{\leavevmode\leaders\hrule height 0.7ex depth \dimexpr0.4pt-0.7ex\hfill\kern0pt}

\usepackage{subcaption}

\begin{document}

\title{Quantum Duality in Electromagnetism and the Fine Structure Constant}

\author{Clay C\'{o}rdova}
\affiliation{Enrico Fermi Institute and Kadanoff Center for Theoretical Physics, University of Chicago}
\author{Kantaro Ohmori}
\affiliation{Faculty of Science, University of Tokyo}

\date{July 24, 2023}

\begin{abstract}

\noindent

We describe the interplay between electric-magnetic duality and higher symmetry in Maxwell theory.  When the fine-structure constant is rational, the theory admits non-invertible symmetries which can be realized as composites of electric-magnetic duality and gauging a discrete subgroup of the one-form global symmetry.  These non-invertible symmetries are approximate quantum invariances of the natural world which emerge in the infrared below the mass scale of charged particles. We construct these symmetries explicitly as topological defects and illustrate their action on local and extended operators. We also describe their action on boundary conditions and illustrate some consequences of the symmetry for Hilbert spaces of the theory defined in finite volume. 

\end{abstract}

\maketitle

\makeatletter
\def\l@subsubsection#1#2{}
\makeatother

\tableofcontents 


\section{Introduction}

Duality is a classical feature of electromagnetism.  In vacuum Maxwell's equations read:
\begin{equation}\label{maxnr}
    \vec{\nabla} \cdot \vec{E}=0~,~ \vec{\nabla} \cdot \vec{B}=0~,~\vec{\nabla} \times \vec{E}=-\frac{\partial{\vec{B}}}{\partial t}~,~\vec{\nabla} \times \vec{B}=\frac{\partial{\vec{E}}}{\partial t}~,
\end{equation}
and are famously invariant under the transformation:
\begin{equation}\label{snrtrans}
    \vec{E}\mapsto\vec{B}~,~~~\vec{B}\mapsto -\vec{E}~.
\end{equation}
This duality survives the inclusion of sources provided that both electric and magnetic charges are allowed and exchange under duality.  
This classical duality transformation holds independent of the value of the electric coupling $e$ which controls the strength of the Coulomb force.

At the quantum level, duality is a more subtle notion.  Adopting standard relativistic notation, Maxwell theory is defined by a path integral over $U(1)$ gauge fields $A$.  The Euclidean action defining the weight in the path integral is:\footnote{In our conventions, the flux $F/2\pi=dA/2\pi$ is integrally quantized,  the gauge field $\frac{1}{e}A$ is canonically normalized, and the $\theta$-angle has periodicity $2\pi$ on spin manifolds.}
\begin{equation}\label{maxact}
S=\frac{1}{2e^{2}}\int F \wedge *F - \frac{i \theta}{8\pi^2}\int F \wedge F~.
\end{equation} 
The $\theta$-angle is a new parameter that controls the weight of topologically non-trivial bundles in the partition function, and below, we simply take $\theta=0.$  The coupling $e$ appears only as an overall constant in the action and hence drops out of the equations of motion 
\begin{equation}
    d *F =0~,
\end{equation}
which encapsulates \eqref{maxnr}.  

The fact that the equations of motion are independent of the coupling can also be understood as follows: at the classical level, the coupling $e$ can be absorbed by rescaling the gauge field $A$ and hence is not meaningful.  In other words, classically there is no preferred unit of electric charge.  By contrast, in the quantum theory Dirac quantization implies that particles carry integer multiples of a basic quantum of electric charge and hence makes the choice of $e$ physical.  Moreover, since magnetic and electric charges are quantized in inverse units, the force between elementary charges is in general not invariant under duality unless the coupling also transforms.  An important consequence of these comments is that quantum mechanically, duality is not in general a property of a fixed theory, but rather is an equivalence between versions of Maxwell theory with different couplings.  

Our purpose in this paper is to explore this fact and to discuss the special circumstances under which duality can be viewed as a symmetry of a fixed quantum theory.  One well known circumstance under which this occurs is when the coupling is tuned to a self-dual value so that the forces between the basic quanta agree.  More generally, below we will show that when $e$ satisfies the rationality condition:
\begin{equation}\label{rational}
    \frac{e^{2}}{2\pi}=\frac{N_{m}}{N_{e}}~,~~~N_{i}\in \mathbb{N}~,~~~\gcd(N_{e},N_{m})=1~,
\end{equation}
Maxwell theory enjoys an exact duality symmetry at the quantum level.  Our result generalizes previous analysis of electric-magnetic duality in \cite{Gaiotto:2008ak, Kapustin:2009av} and duality defects in \cite{Choi:2021kmx, Choi:2022zal, Apte:2022xtu} which discussed the symmetry associated to the special case $N_{m}=1.$  In particular, like the latter analysis, the symmetries we construct are in general non-invertible, and thus while they commute with the Hamiltonian, they are not represented by unitary operators acting on Hilbert space. The fact that non-invertible symmetries exist in Maxwell theory at more general rational couplings like \eqref{rational} was first pointed out in \cite{Niro:2022ctq}.  Our analysis below extends these observations to reveal the particularly simple nature of the associated symmetry defects and to describe their properties, like fusion rules, and their physical consequences for boundary conditions and Hilbert spaces. Our work follows on a variety of recent analysis describing non-invertible symmetry in field theory, particularly in 3+1 dimensions, e.g.~\cite{Tachikawa:2017gyf,Choi:2021kmx,Wang:2021vki,Hayashi:2022oxp,Anosova:2022yqx,Roumpedakis:2022aik,Bhardwaj:2022yxj,Arias-Tamargo:2022nlf,Hayashi:2022fkw,Choi:2022zal,Kaidi:2022uux,Choi:2022jqy,Cordova:2022ieu,Antinucci:2022eat,Choi:2022rfe,Bhardwaj:2022lsg,Bartsch:2022mpm,Freed:2022qnc,Kaidi:2022cpf,Chen:2022cyw,Karasik:2022kkq,Cordova:2022fhg,GarciaEtxebarria:2022jky,Choi:2022fgx,Yokokura:2022alv,Bhardwaj:2022kot,Bhardwaj:2022maz,Bartsch:2022ytj,Hsin:2022heo,Das:2022fho,Apte:2022xtu,Kaidi:2023maf,Brennan:2023kpw,Putrov:2023jqi,Koide:2023rqd,Bhardwaj:2023wzd,Bhardwaj:2023ayw,Bartsch:2023wvv,Damia:2023ses,Copetti:2023mcq,Argurio:2023lwl,vanBeest:2023dbu,Chen:2023czk}.

Notice that the critical values identified in \eqref{rational} correspond to rational fine structure constant.  These are a dense, but measure zero set, in the space of all allowed couplings.  In nature, the fine structure constant runs logarithmically due to the presence of massive charged leptons and quarks.  In the infrared, far below the mass scale of these charged particles the value stabilizes at:
\begin{equation}
    \frac{e_{\text{IR}}^{2}}{2\pi}\approx \frac{2}{137.03599908}~.
    \label{eq: real alpha}
\end{equation}
We can view this decimal approximation as a sequence of rationals with increasing precision:
\begin{equation}\label{approx}
    \frac{N_{m}^{0}}{N_{e}^{0}}=\frac{2}{137}~,~~\frac{N_{m}^{1}}{N_{e}^{1}}=\frac{25}{1713}~,~~\frac{N_{m}^{2}}{N_{e}^{2}}=\frac{500}{34259}~, \cdots~.
\end{equation}
Truncating to a fixed precision, we conclude that the symmetries we identify can be viewed as approximate symmetries of the natural world which emerge at low energy in the quantum realm of photons.\footnote{It is interesting to ask how the $\theta$-angle in Maxwell theory might modify this observation.  In the Standard Model, we do not know the value of $\theta$ and the details of the non-invertible symmetry in general depend on $\theta.$ Here for simplicity, we will assume that $\theta$ vanishes. It would be interesting to explore this issue in more detail in future work.}

In addition to simply identifying these special values of the coupling, we will also describe the action of the corresponding symmetries on operators.  To carry this out, we construct topological defects that represent the symmetry action which generalize those constructed in \cite{Choi:2021kmx, Choi:2022zal}. Explicitly, these defects are $\mathbb{Z}_{N_{m}}$ topological gauge theories which couple to the bulk gauge fields as in a topological order.  The defects actions we construct may be viewed as simplifications of those derived in \cite{Niro:2022ctq, email}.  For instance, in the leading approximation mentioned in \eqref{approx}, the approximate quantum symmetry of Maxwell theory is mediated by coupling to a $\mathbb{Z}_{2}$ topological field theory, i.e.\ a toric code state.  

We use these defect actions to describe how the symmetries act on boundary conditions, generalizing the analysis of electric-magnetic duality in \cite{Gaiotto:2008ak, Kapustin:2009av} and analogous previous studies \cite{Koide:2023rqd, Choi:2023xjw}.  To illustrate this action in detail we consider a quasi-realistic setup described by Maxwell theory in a toroidal cavity: a spatial solid torus with specified boundary conditions. At the special rational coupling \eqref{rational}, the energy levels and degeneracies in this Hilbert space are invariant under exchanging electric conducting boundary conditions (Dirichlet), with magnetic conducting boundary conditions (Neumann) provided that we also couple to a $\mathbb{Z}_{N_{m}}$ topological sector. 

In our universe, free magnetic charges have not been observed and certainly do not propagate at low energies, making direct experimental verification of this equivalence difficult.  Nevertheless it may perhaps be possible to design materials that simulate magnetic conductors in a range of frequencies or energies.  In such a hypothetical material, the matching of energy levels arising from these approximate symmetries of our universe might be tested in future experiments.

\section{Electric-Magnetic Duality}

Let us first review the standard derivation of electric-magnetic duality in the quantum setup. We refer to this duality operation as $\mathbb{S},$ to distinguish it from other notions of duality defined below. 

In the path integral, we treat $F$ as the fundamental variable instead of $A$.  The Bianchi identity, $dF=0,$ is then achieved by introducing a Lagrange multiplier gauge field $\widetilde{A}$ together with a coupling
\begin{equation}
  S \supset  \frac{i}{2\pi}\int d\widetilde{A} \wedge F ~.
\end{equation}
To see that this is the correct normalization, note that in the presence of a charge $m$ magnetic monopole inserted on a world line $\gamma$ the term above reduces as:
\begin{equation}
    \frac{i}{2\pi}\int d\widetilde{A} \wedge F \longrightarrow i m \int_{\gamma}\widetilde{A}~,
\end{equation}
so that $\widetilde{A}$ couples to the magnetic charge in the exact same manner as $A$ couples to electric charge.  Now integrate out $F$ to obtain the duality relation between the field strength $F$ and its dual $\widetilde{F}=d\widetilde{A}$:
\begin{equation}\label{dualityrel}
  \mathbb{S}(F)= \widetilde{F} = \frac{2\pi i}{e^{2}}*F~.
\end{equation}
Using this relationship we can write the action solely in terms of dual variables with a dual coupling $\widetilde{e}$:
\begin{equation}\label{dualityact}
    S=\frac{1}{2\widetilde{e}^{2}}\int \widetilde{F} \wedge *\widetilde{F}~, \hspace{.2in} \frac{e^{2}}{2\pi}=\frac{2\pi}{\widetilde{e}^{\phantom{.}2}}~.
\end{equation}
Thus as is familiar, electric-magnetic duality inverts the coupling constant and hence relates a theory with coupling $e$ to a dual theory with coupling $\mathbb{S}(e)=\widetilde{e}.$

Since the local operators in Maxwell theory are generated by field strengths, \eqref{dualityrel} yields the general rule for their transformation under $\mathbb{S}$.  Meanwhile to see the action on extended operators it is useful to organize the action of duality via the one-form symmetry.  Maxwell theory has a $U(1)^{(1)}_{e}\times U(1)^{(1)}_{m}$ symmetry \cite{Gaiotto:2014kfa} with charges generated by surface operators:
\begin{itemize}
    \item Electric symmetry $U(1)^{(1)}_{e}$: The charge is the surface operator $Q_{E}=\frac{ 1}{ie^{2}}\int * F$.  The charged objects are Wilson lines $\exp(i\ell\int A)$ for $\ell\in \mathbb{Z}$.
\item Magnetic symmetry $U(1)^{(1)}_{m}$: The charge is the surface operator $Q_{M}=\frac{ 1}{2\pi}\int  F$.  The charged objects are 't Hooft lines.
\end{itemize}

We can also describe the electric one-form symmetry generator in more physical terms as follows.  We work locally in $\mathbb{R}^{4}$ with cylindrical coordinates $(r,\theta,z,t).$  Consider a solenoid of radius $R$ centered at $r=0$ and extending spatially along $z$.  Inside the solenoid there is a constant magnetic field of magnitude $B$ in the $z$ direction, while outside the flux vanishes. This is described by a gauge field
\begin{equation}\label{1forms}
    A= \begin{cases}
    \left(\pi r^{2}B\right)\frac{d\theta}{2\pi}~, & r <R~, \\
    &\\
    \left(\pi R^{2}B\right)\frac{d\theta}{2\pi} ~, & r>R~.
    \end{cases}
\end{equation}
Consider the limit $R\rightarrow 0,$ and $B\rightarrow \infty$ with the total magnetic flux $\pi R^{2}B \equiv \alpha$ held fixed.  Then the field strength everywhere vanishes (except formally at $r=0$), and the above defines a surface defect localized at $r=0$ and extended in $z$ and $t$. This defect is detectable through its fixed Aharanov-Bohm phase.  This is precisely the (electric) one-form symmetry defect which is characterized by the fact that Wilson lines encircling it have a fixed value $\alpha$ (defined modulo $2\pi$).  

A similar electric flux confined in an infinitesimal region defines the magnetic one-form symmetry generators. Such a configuration can be approximated by an array of tiny capacitors. Then, under electric-magnetic duality $\mathbb{S},$ these two classes of generators, or defects, are exchanged consistent with \eqref{smap}. 

From the relation \eqref{dualityrel} we can see that the one-form symmetries \eqref{1forms} map under duality as:
\begin{equation}\label{smap}
   \mathbb{S}(Q_{E})= \widetilde{Q}_{E}=Q_{M}~, ~~~\mathbb{S}(Q_{M})= \widetilde{Q}_{M}=-Q_{E}~,
\end{equation}
which in particular also determines the action on the charged Wilson and 't Hooft lines.  We also note that duality squares to charge conjugation.

\subsection{Interfaces and Boundary Conditions}

The previous discussion explains how electric-magnetic duality acts on point, line, and surface operators.  Here we discuss how it acts on operators of codimension one, i.e.\ interfaces and boundaries.  Our analysis closely follows \cite{Kapustin:2009av}.  Let us begin by enumerating several natural boundary conditions.

\emph{Perfect electric conductor.}  The field strength restricts to be trivial on the boundary:
    \begin{equation}\label{peccond}
        F|_{\partial M}=0~.
    \end{equation}
    Thus, the electric field is perpendicular to the spatial boundary, while the magnetic field is parallel to it.  Working in a gauge where $A_{\hat{n}}$ is zero for $\hat{n}$ a normal direction to $\partial M$, and assuming that the boundary topology is trivial, one can equivalently view \eqref{peccond} as defining Dirichlet boundary conditions $A|_{\partial M}=0$.\footnote{More generally, when the topology of $\partial M$ is not trivial boundary condition defined by \eqref{peccond} has moduli corresponding to flat connection on $\partial M$.  Ignoring torsion, these are parameterized by a torus  $H^{1}(\partial M, \mathbb{R})/2\pi H^{1}(\partial M, \mathbb{Z}).$} 
    
    It is convenient to model the boundary condition by introducing a new $3d$ dynamical $U(1)$ gauge field $c$ whose equation of motion enforces \eqref{peccond}.  The required boundary action is a mixed Chern-Simons term:
\begin{equation}\label{PEC}
   S_{b}= \frac{i}{2\pi}\int_{\partial M} c \wedge dA~.
\end{equation}
Taking into account the boundary variation of the bulk action we find that in addition to \eqref{peccond} 
\begin{equation}
    \frac{i}{e^{2}}*F|_{\partial M} = \frac{dc}{2\pi}~.
\end{equation}
Thus the electric one-form symmetry current is identified on the boundary with the flux of $c$.

\emph{Perfect magnetic conductor.}  In this case the dual field strength restricts to be trivial on the boundary:
 \begin{equation}\label{pmccond}
        *F|_{\partial M}=0~.
    \end{equation}
    Thus the magnetic field is perpendicular to the spatial boundary, while the electric field is parallel to it. Again assuming the gauge $A_{\hat{n}}=0$ and trivial boundary topology, \eqref{pmccond} yields Neumann boundary conditions $\partial_{\hat{n}}A|_{\partial M}=0.$ Note that since the boundary variation of the bulk action is trivialized when \eqref{pmccond} holds,  we can enforce this boundary condition using only the bulk action without any additional boundary fields.   

\emph{General Boundary Conditions.} More generally, we may consider a boundary condition described by coupling $n$ boundary fields $c_{\alpha}$ to the bulk.  Such a boundary is described by a symmetric integral level matrix $k_{\alpha\beta},$ an integral vector $v_{\alpha},$ and an integer $p$ with boundary action:
\begin{equation}\label{Sbgen}
  S_{b}=  \frac{i k_{\alpha\beta}}{4\pi} \int_{\partial M}c_{\alpha}\wedge dc_{\beta}+\frac{iv_{\alpha}}{2\pi}\int_{\partial M}c_{\alpha}\wedge dA+\frac{ip}{4\pi}\int_{\partial M}A\wedge dA~.
\end{equation}
For such a boundary condition, neither one-form symmetry is trivialized on the boundary, but instead they are related to the fluxes of $c_{\alpha}$ as:
\begin{eqnarray}
    k_{\alpha\beta}\frac{dc_{\beta}}{2\pi}+v_{\alpha}\frac{F}{2\pi}|_{\partial M}&=&0~. \\
    \frac{1}{ie^2}*F|_{\partial M}+v_{\alpha}\frac{dc_{\alpha}}{2\pi}+p\frac{F}{2\pi}|_{\partial M}&=&0~.
\end{eqnarray}
Assuming $k_{\alpha\beta}$ is non-degenerate one may formally integrate out the boundary gauge fields $c_{\alpha}$ resulting in a Chern-Simons term for $A$ with a fractional level:
\begin{equation}
    (p-v_{\alpha}(k^{-1})_{\alpha\beta}v_{\beta})\frac{i}{4\pi}\int_{\partial M}A\wedge dA~.
\end{equation}
The true quantum consistent theory is described by \eqref{Sbgen}.

To deduce the action of electric-magnetic duality on the boundary conditions described above, we describe the $\mathbb{S}$ duality itself as a topological interface connecting a theory with gauge field $A_{L}$ and coupling $e$ with a theory with gauge field $A_{R}$ and coupling $\tilde{e}$ defined in \eqref{dualityact}.  The interface action is simply \cite{Gaiotto:2008ak, Kapustin:2009av}:
\begin{equation}\label{Sinterface}
    S=\frac{i}{2\pi}\int_{W} A_{L} \wedge d A_{R}~,
\end{equation}
where $W$ indicates the location of the wall.  The full bulk and boundary action then consists of Maxwell actions \eqref{maxact} for $A_{L}$ and $A_{R}$ together with the interface coupling \eqref{Sinterface}.  

The boundary term in the equations of motion then gives rise to a continuity equation for the bulk gauge fields along the wall:\footnote{In \eqref{continuitys} $*$ denotes the four-dimensional Hodge star operator and thus the stated equations are independent. }
\begin{equation}\label{continuitys}
    \left(\frac{2\pi i}{e^{2}}*F_{L}-F_{R}\right)|_{W}=0~,~~~\left(F_{L}+\frac{2\pi i}{\tilde{e}^{2}}*F_{R}\right)|_{W}=0~,
\end{equation}
thus reproducing \eqref{dualityrel}.

\begin{figure}
    \centering
    \includegraphics[width=.8\linewidth]{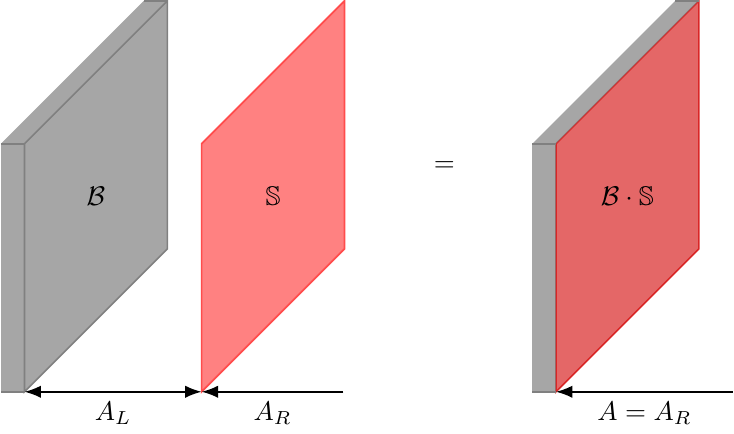}
    \caption{The $\mathbb{S}$ action on a boundary condition $\mathcal{B}$.  The symmetry defect (red) collides with  $\mathcal{B}$ to make a new boundary condition $\mathcal{B}\cdot \mathbb{S},$ with $A_{L}$ a new boundary field.}
    \label{fig: defect boundary}
\end{figure}

The interface \eqref{Sinterface} allows us to determine the action of $\mathbb{S}$ duality on boundary conditions.  We place the defect \eqref{Sinterface} near the boundary and parallel to it. In the limit when the defect collides with the boundary, one of the bulk fields, say $A_{L}$ becomes a new boundary gauge field $a$, while $A_{R} \rightarrow A$, becomes the new bulk gauge field.  See Figure \ref{fig: defect boundary}.

As described in \cite{Kapustin:2009av}, this provides an elegant derivation that Dirichlet and Neumann boundary conditions are exchanged under $\mathbb{S}$-duality. Indeed, acting for instance on a perfect magnetic conductor with trivial boundary action, the duality defect \eqref{Sinterface} yields 
\begin{equation}\label{PECs}
   S_{b}= \frac{i}{2\pi}\int_{\partial M} a \wedge dA~,
\end{equation}
exactly producing the perfect electric conductor.  Similarly, the dual of the perfect electric conductor is a perfect magnetic conductor. 

\subsection{$\mathbb{S}$-Duality and Hilbert Spaces}\label{sechilb}

To illustrate the concepts above, it is instructive to consider duality in the context of a Hilbert space.  An especially simple physical set up is a toroidal cavity, i.e.\ space is a solid torus with boundary conditions, which we take here to either be a perfect electric conductor, i.e. Dirichlet, or a perfect magnetic conductor, i.e. Neumann.  Throughout, we work in temporal gauge $A_{t}=0$, and take coordinates $(r,\theta, \phi)$ on the solid torus.  The angles are periodic and $r$ is bounded:
\begin{equation}
    \theta \sim \theta+2\pi~, ~~~\phi \sim \phi +2\pi~, ~~~ 0\leq r \leq R~.
\end{equation}
The circle parameterized by $\theta$ has radius $r$ and is contractible in the interior of the solid torus.  The circle parameterized by $\phi$ is non-contractible and has length $L$.  The metric on space is:
\begin{equation}
    ds^{2}=L^{2}d\phi^{2}+dr^{2}+r^{2}d\theta^{2}~.
\end{equation}

\subsubsection{Dirichlet}

We consider a subset of modes defining a sub-Hilbert space $\mathcal{H}$ where the only non-vanishing components of the gauge field are $A_{\theta}.$  
In general this gives rise to a field strength $F_{r\theta}=\partial_{r}A_{\theta}.$ Let us determine its profile.  The equation of motion and boundary conditions are:
\begin{equation}
   \partial_{r}\left(\frac{\partial_{r}A_{\theta}}{r}\right)=0~,~~~ A_{\theta}|_{r=R}=0~,
\end{equation}
which is solved by:
\begin{equation}\label{d1sol}
    A_{\theta}=\beta(r^{2}-R^{2})~,
\end{equation}
with $\beta$ constant.  We must also impose that the solution is regular as $r\rightarrow 0$. To do so, we note that as $r$ approaches the origin, the circle parameterized by $\theta$ shrinks and hence the holonomy around this circle must approach unity:
\begin{equation}
    \lim_{r\rightarrow 0}\exp\left(i\oint A\right)=1~.
\end{equation}
This condition quantizes the parameter $\beta$ appearing in \eqref{d1sol} leading to solutions:
\begin{equation}\label{d2sol}
    A_{\theta}=n\left(\frac{r^{2}}{R^{2}}-1\right)~.
\end{equation}
As a consistency check, we notice that the flux through the disc $(r,\theta)$ at constant $\phi$ is properly quantized.\footnote{On a closed spacetime manifold $M$ the flux is quantized as $F/2\pi\in H^{2}(M,\mathbb{Z})$.  On a manifold with boundary with Dirichlet boundary conditions, the appropriate fluxes are modified to relative cohomology classes $F/2\pi\in H^{2}(M,\partial M,\mathbb{Z}).$ }

It is straightforward to use the action \eqref{maxact} to evaluate the energy of these states:
\begin{equation}\label{end}
    E_{n}=\left(\frac{4\pi^{2}L}{e^{2}R^{2}}\right)n^{2}~.
\end{equation}

\subsubsection{Neumann}

The dual problem involves a toroidal cavity with Neumann boundary conditions.  The dual field configurations to \eqref{d2sol} are those with a purely electric field $F_{t\phi}$.  Restricted to these modes, the action takes the form 
\begin{equation}\label{n1sol}
   S= \frac{\pi^{2} R^{2}}{L\tilde{e}^{2}}\int \dot{A}_{\phi}^{2} ~ dt~.
\end{equation}
The Neumann boundary conditions are automatic since $A_{\phi}$ is independent of $r$.  Since $A_{\phi}$ is a gauge field it is naturally a periodic variable with unit periodicity.  Thus \eqref{n1sol} is the action for a quantum particle on a circle.  The energy spectrum is obtained by quantizing the momentum:
\begin{equation}
P=\frac{2\pi^{2}R^{2}}{L\tilde{e}^{2}}\dot{A}_{\phi} =2\pi n~,~~~n\in \mathbb{Z}~.
\end{equation}
The resulting energies are then:
\begin{equation}
    E_{n}=\left(\frac{L \tilde{e}^{2}}{R^{2}}\right)n^{2}~,
\end{equation}
which agree with \eqref{end} precisely when the couplings $e$ and $\tilde{e}$ are related by duality \eqref{dualityrel}.

\section{Dualities, Gauging, and Symmetries }

One aspect of Maxwell theory and duality which is made explicit by the Hilbert space calculations of the previous section is the dependence on the coupling. Indeed, the coupling is an overall factor in the action and hence at the classical level does not enter the equations of motion. By contrast in the quantum theory, the energy levels in general depend explicitly on $e$.  As emphasized above, this means that electric-magnetic duality is not a symmetry of the quantum theory, but is instead a map between one theory with coupling $e$ and another theory with coupling $\tilde{e}.$

However, at a special value of the coupling, $\mathbb{S}$ can be viewed as a map acting on a fixed theory.  This occurs when
\begin{equation}\label{especial}
    e=\tilde{e}=\sqrt{2\pi}~.
\end{equation}
Here, for example the energy levels of the theory quantized with perfect electric conducting boundary conditions are identical to those with perfect magnetic conducting boundary conditions. 

More abstractly, at the coupling \eqref{especial}, $\mathbb{S}$ defines an internal discrete 0-form global symmetry. Inspecting the transformation \eqref{snrtrans}, we see that $\mathbb{S}$ squares to charge conjugation and so defines the generator of a $\mathbb{Z}_{4}$ symmetry.  This symmetry is characterized by its associated codimension one topological defect given by \eqref{Sinterface}, where now $A_{L}$ and $A_{R}$ are interpreted as gauge fields in the same Maxwell theory at coupling \eqref{especial}.

\subsection{Self-Duality at Rational Coupling}

More generally, there are versions of self-duality that exist for any rational coupling:
\begin{equation}\label{rational2}
    \frac{e^{2}}{2\pi}=\frac{N_{m}}{N_{e}}~,~~~N_{i}\in \mathbb{N}~,~~~\gcd(N_{e},N_{m})=1~.
\end{equation}
These can be constructed by gauging a discrete subgroup $\mathbb{Z}_{N_{e}}^{(1)}\times \mathbb{Z}_{N_{m}}^{(1)}$ of the $U(1)^{(1)}_{e}\times U(1)^{(1)}_{m}$ one-form symmetry and then combining with electric-magnetic duality. The case of $\mathbb{S}$ discussed above in \eqref{especial} corresponds to $N_{e}=N_{m}=1,$ and the case of $N_{m}=1$ was described in \cite{Choi:2021kmx}.  Meanwhile the existence to non-invertible symmetry at general rational coupling was first observed in \cite{Niro:2022ctq}.

In order to demonstrate the existence of these self duality symmetries let us first recall how to couple Maxwell theory to background gauge fields for the one-form symmetry.  The action takes the form:
\begin{equation}\label{backs}
    S=\frac{1}{2e^{2}}\int (F-B_{e}) \wedge * (F-B_{e})+\frac{i}{2\pi}\int (F-B_{e}) \wedge B_{m}~,
\end{equation}
where $B_{i}$ is a two-form background field subject to the gauge redundancy:\footnote{Under the gauge transformation of the electric one-form background $B_{e},$ the dynamical gauge field $A$ also shifts since the photon is the Goldstone mode of the one-form symmetry.}
\begin{equation}\label{backgt}
    B_{i}\sim B_{i}+d\Lambda_{i}~.
\end{equation}
In the action \eqref{backs}, the background gauge fields couple linearly to the one-form symmetry currents discussed above \eqref{smap}.  The choice of counterterms, i.e. terms depending only on the background fields is selected for later convenience.  

Observe that the action \eqref{backs} is not exactly gauge invariant under \eqref{backgt}.  Rather it transforms as:
\begin{equation}
    \delta S=\frac{1}{2\pi i}\int B_{e}\wedge d\Lambda_{m}~.
\end{equation}
This is the appropriate transformation law for an 't Hooft anomaly.  It may be cancelled by inflow from a 5d classical action:
\begin{equation}\label{anomaly}
    \mathcal{A}=\frac{i}{2\pi}\int dB_{e}\wedge B_{m}~.
\end{equation}
Here, the integral above defining $\mathcal{A}$ is taken over a five manifold with boundary the physical spacetime, and the anomaly should be interpreted as an exponentiated action, i.e.\ $\exp(\mathcal{A})$ must be well defined, but $\mathcal{A}$ need only be defined up to shifts by $2\pi i \mathbb{Z}$.  The presence of this anomaly means that in general it is not consistent to simultaneously gauge the electric and magnetic one-form symmetries.  

In our case however, we are interested in gauging a discrete $\mathbb{Z}_{N_{e}}^{(1)}\times \mathbb{Z}_{N_{m}}^{(1)}$ subgroup of the total $U(1)_{e}^{(1)}\times U(1)_{m}^{(1)}$ one-form symmetry with $\gcd(N_{e},N_{m})=1$.  Thus we aim to show that the anomaly \eqref{anomaly} trivializes in this case. 
Intuitively, one can see this triviality from the  mathematics of background fields, i.e.\ from the fact that there is no meaningful product between $\mathbb{Z}_n$ and $\mathbb{Z}_m$ cochains when $\gcd(n,m)=1$.  Below, we explicitly demonstrate this triviality and along the way construct the action where $\mathbb{Z}_{N_{e}}^{(1)}\times \mathbb{Z}_{N_{m}}^{(1)}$ can be gauged.

To carry out this restriction we fix the backgrounds to be flat $dB_{i}=0$ and restrict the holonomies appropriately as
\begin{equation}
    \hat{C}_{i}=\frac{N_{i}}{2\pi}B_{i}\in C^{2}(M,\mathbb{Z})~,
\end{equation}
which are $\mathbb{Z}$-uplifts of $\mathbb{Z}_{N_i}$ cycles $C_i \in H^2(M,\mathbb{Z}_{N_i})$.
The anomaly \eqref{anomaly} simplifies to:
\begin{equation}\label{discreteanom}
     \mathcal{A}=\frac{2\pi i}{N_{m}}\int \beta(\hat{C}_{e}) \cup \hat{C}_{m}~,
\end{equation}
where above, $\cup$ is the cup-product among $\mathbb{Z}$-cochains, and $\beta(C_{e})$ denotes the Bockstein:
\begin{equation}\label{bockdef}
    \beta(C_{i})=\frac{\delta \hat{C}_{i}}{N_{i}} \in H^{3}(M,\mathbb{Z})~.
\end{equation}
The Bockstein is a discrete analog of the curvature of $C_{i}$.  From \eqref{bockdef}, we see that it is necessarily a torsion class: $N_{i}\beta(C_{i})=\delta \hat{C}_i.$

Superficially, the anomaly \eqref{discreteanom} appears to be non-vanishing.  To investigate it, we now use the fact that $\gcd(N_{e},N_{m})=1$ to find integers $x,y$ solving
\begin{equation}
    x N_{e}+yN_{m}=1~.
\end{equation}
The anomaly formula \eqref{discreteanom} is invariant, up to $2\pi i$ times integers, under shifts in $\hat{C}_{m}$ by multiples of $N_{m}$ times any cochain.  Thus we have:
\begin{eqnarray}
    \mathcal{A} & = & \frac{2\pi i}{N_{m}}\int  \left(1-yN_{m}\right)\beta(C_{e}) \cup \hat{C}_{m}~,\nonumber \\
    & = & \frac{2\pi i}{N_{m}}\int \left(x N_{e}\right)\beta(C_{e})  \cup \hat{C}_{m}~. 
\end{eqnarray}
Now we use the fact mentioned below \eqref{bockdef}, namely that the Bockstein is torsion to simplify, along with integration by parts to simplify:
\begin{equation}\label{anomvanish}
    \mathcal{A}  =  \frac{2\pi i x}{N_{m}} \int \delta \hat{C}_{e} \cup \hat C_{m}=- \frac{2\pi i x}{N_{m}} \int  \hat{C}_{e} \cup \delta\hat C_{m} \in 2\pi i \mathbb{Z}~,
\end{equation}
where in the final step of \eqref{anomvanish} we have used the assumption that the $N_m$ reduction of $\hat{C}_m$ is closed.
Thus we conclude for $\gcd(N_{e},N_{m})=1$ we may gauge the $\mathbb{Z}_{N_{e}}^{(1)}\times \mathbb{Z}_{N_{m}}^{(1)}$ one-form global symmetry.  

Having deduced that the gauging is consistent, we now elucidate its effect on Maxwell theory. We will see that the result is again Maxwell theory but with modified coupling.  One way to interpret \eqref{anomvanish} is that since $N_{e}$ and $N_{m}$ are coprime, we can freely rescale $\hat{C}_{m}$ by $x N_{e}$, and doing so makes the anomaly vanish.  If we carry this out in the action \eqref{backs} we find:
\begin{equation}\label{dynam}
     S=\frac{2\pi^{2}}{e^{2}}\int \left(\frac{F}{2\pi}-\frac{\hat{C}_{e}}{N_{e}}\right)^{2} +\frac{2\pi i x N_{e}}{N_{m}}\int \left(\frac{F}{2\pi}-\frac{\hat{C}_{e}}{N_{e}}\right) \cup \hat{C}_{m}~.
\end{equation}
Above, the fields $\hat{C}_{i}$ are now dynamical and our aim is to sum over them.  

It is straightforward to interpret the above as the action of a new Maxwell theory with modified coupling constants.  Indeed, ignoring the coupling to $\hat{C}_{m}$ the $\hat{C}_{e}$ gauge field effectively allows fractional fluxes:
\begin{equation}
    \int_{\Sigma} \left(\frac{F}{2\pi}-\frac{\hat{C}_{e}}{N_{e}}\right)=\frac{n}{N_{e}}~,~~~n\in \mathbb{Z}~,
\end{equation}
where $\Sigma$ is any two-cycle.  The sum over $\hat{C}_{m}$ then restricts the numerator $n$ above to be a multiple of $N_{m}$.  
In summary we may view \eqref{dynam} as a theory of a $U(1)$ gauge field whose fluxes are quantized to be integer multiples of the fraction $N_{m}/N_{e}$. Or rescalling to a canonically normalized gauge field, we find the action of gauging on the coupling constant:
\begin{equation}\label{gaugecoupling}
    \text{gauging}~\mathbb{Z}_{N_{e}}^{(1)}\times \mathbb{Z}_{N_{m}}^{(1)}:~e\mapsto \left(\frac{N_{e}}{N_{m}}\right)e~.
\end{equation}

Finally, we can combine our analysis above with the discussion of $\mathbb{S}$ to derive the claimed self-duality symmetry. Specifically, if we compose \eqref{gaugecoupling} with the action of $\mathbb{S}$ in \eqref{dualityact}, we see that the theory is at a fixed point exactly when the coupling satisfies the rationality constraint \eqref{rational}.  We denote the resulting symmetry by $\mathcal{D}_{N_{e},N_{m}}:$
\begin{equation}
    \mathcal{D}_{N_{e},N_{m}} \equiv \mathbb{S} \circ  \text{gauging}~\mathbb{Z}_{N_{e}}^{(1)}\times \mathbb{Z}_{N_{m}}^{(1)}~.
\end{equation}
Below we derive the properties of this symmetry.

\subsection{Interfaces}

To implement the symmetry $\mathcal{D}_{N_{e},N_{m}}$ explicitly at the level of operators we must construct a topological interface.  This is codimension one defect generalizing the $\mathbb{S}$ defect described in \eqref{Sinterface} and the case of $N_{m}=1$ constructed in \cite{Choi:2021kmx}.  For the case of general rational coupling the defects we construct may be viewed as simplifications of those derived in \cite{Niro:2022ctq, email}.

Its topological nature reflects the fact that it is a symmetry of this theory.  In general, we expect the interface to be described by dynamical Chern-Simons gauge fields which couple the to bulk physics through the restrictions $A_{L}$ and $A_{R}$ of the bulk electromagnetic gauge fields on the left and right of the defect.

Let us first describe the answer intuitively before giving a more rigorous construction.  We consider the interface action to be simply:
\begin{equation}\label{intuitived}
    S_{\mathcal{D}_{N_{e},N_{m}}} \sim \frac{i}{2\pi}\left(\frac{N_{e}}{N_{m}}\right)\int_{W} A_{L} \wedge d A_{R}~,
\end{equation}
where $W$ indicates the location of the wall.  The full bulk and boundary action then consists of Maxwell actions \eqref{maxact} for $A_{L}$ and $A_{R}$ together with the interface coupling \eqref{intuitived}.  The boundary term in the equations of motion then gives rise to a continuity equation for the bulk gauge fields along the wall:\footnote{In \eqref{continuity} $*$ denotes the four-dimensional Hodge star operator and thus the stated equations are independent. }
\begin{equation}\label{continuity}
    (*F_{L}+iF_{R})|_{W}=0~,~~~(F_{L}+i*F_{R})|_{W}=0~.
\end{equation}
To see that these continuity equations define a topological defect, we must check that the energy-momentum tensor is continuous when passing through the interface.  In general, we have:
\begin{equation}\label{tem}
    T_{\mu\nu}=\frac{1}{e^{2}}\left(F_{\mu\alpha}F_{\alpha\nu}+\frac{1}{4}g_{\mu\nu}F_{\alpha\beta}F^{\alpha\beta}\right)~.
\end{equation}
It is instructive to reduce this into components. Let $i,j=1,2,3$ be coordinates along $W$ and $n$ a normal direction to $W$.  We assume the coordinates are orthonormal with orientation chosen so that $\epsilon_{n123}=1.$  The field strength decomposes into three-vectors $X$ and $Y$ along $W$ with components:
\begin{equation}
    F_{ni}=X_{i}~, ~~~Y_{i}=-\frac{1}{2}\epsilon_{ijk}F_{jk}~.
\end{equation}
Then the energy momentum tensor is:
\begin{eqnarray}\label{tnonrel}
    T_{nn}=\frac{1}{2e^{2}}\left(X^{2}-Y^{2}\right)~,~~~T_{ni}=\frac{1}{e^{2}}\epsilon_{ijk}X_{j}Y_{k}~, \nonumber\\
    T_{ij}=\frac{1}{e^{2}}\left(Y_{\ell}Y_{m}-X_{\ell}X_{m}\right)\left(\delta_{\ell i}\delta_{m j}-\frac{1}{2}\delta_{\ell m}\delta_{ij}\right)~.
\end{eqnarray}
The boundary condition \eqref{continuity} reads:
\begin{equation}
    X_{L}|_{W}=iY_{R}|_{W}~, ~~~Y_{L}|_{W}=iX_{R}|_{W}~.
\end{equation}
Using these equations it is straightforward to check continuity of \eqref{tnonrel} across the wall $W$ and hence verify that \eqref{intuitived} defines a topological defect.

The reason that \eqref{intuitived} does not properly define a topological defect is that the Chern-Simons level appearing on the wall is not quantized when $N_{m}>1.$  To remedy this we must unfold the worldvolume theory \eqref{intuitived} to a Chern-Simons theory with properly quantized levels and additional dynamical fields so that the fractional level arises as an effective response.  

It is straightforward to carry this out by composing a sequence of interfaces which successively implement the gauging operations and then electric-magnetic duality.  The interface that connects two-versions of Maxwell theory which differ by $\mathbb{Z}_{N_{e}}^{(1)}$ gauging has a worldvolume action
\begin{equation}\label{egauge}
    S=\frac{i}{2\pi}\int_{W} a \wedge (N_{e}dA_{L}-dA_{R})~,
\end{equation}
where $a$ indicates a dynamical gauge field that resides only on the wall.  The equation of motion for $a$ imposes a constraint $N_{e}A_{L}|_{W}=A_{R}|_{W}$.  Similarly, the topological wall implementing a $\mathbb{Z}_{N_{m}}^{(1)}$ gauging has a worldvolume action:
\begin{equation}\label{mgauge}
    S=\frac{i}{2\pi}\int_{W} a \wedge (dA_{L}-N_{m}dA_{R})~.
\end{equation}
Finally, the electric-magnetic duality wall is described by \eqref{intuitived} with $N_{e}=N_{m}=1.$  Analogous to the action of $\mathbb{S}$ on boundaries, we can compose these defects by concatentating them in space and interpreting the intermediate external fields as new defect fields.  Carrying this out yields our desired worldvolume theory with action:
\begin{equation}
    S=\frac{i}{2\pi}\int_{W} \hspace{-.05in} a_{1} \wedge (N_{e}dA_{L}-dc)+  a_{2} \wedge (dc-N_{m}db)+b \wedge dA_{R}~.
\end{equation}
Integrating out $c$ enforces $a_{1}=a_{2} \equiv -a$ yielding:
\begin{equation}\label{fulldefect}
   S_{\mathcal{D}_{N_{e},N_{m}}}= \frac{i}{2\pi}\int_{W} N_{m} a \wedge db  - N_{e}a \wedge dA_{L} + b \wedge dA_{R}~.
\end{equation}
The action \eqref{fulldefect} gives a consistent quantum definition of the duality defect $\mathcal{D}_{N_{e},N_{m}}$ which exists at the special value of the coupling \eqref{rational2}.  Note that formally integrating out the dynamical fields $a$ and $b$ and substituting back in indeed gives the anticipated response \eqref{intuitived}.  However even when the bulk fields $A_{i}$ restrict to be trivial on the wall, the defect defined by \eqref{fulldefect} is non-trivial and yields a $\mathbb{Z}_{N_{m}}$ topological gauge theory in $2+1$ dimensions.

\subsection{Fusion Rules}

When $N_{e}$ and $N_{m}$ are not both one, the symmetry we have constructed is non-invertible.  To illustrate this, we examine the fusion $\mathcal{D}_{N_e,N_m}\times \overline{\mathcal{D}}_{N_e,N_m}$ where $\overline{\mathcal{D}}_{N_e,N_m}$ indicates the CPT conjugate of the defect $\mathcal{D}_{N_e,N_m}$ and is defined by placing $\mathcal{D}_{N_e,N_m}$ on a manifold with opposite orientation.   

The action for the composite defect $\mathcal{D}_{N_e,N_m}\times \overline{\mathcal{D}}_{N_e,N_m}$ is defined as usual by concatenating the actions and incorporating the middle region as a new defect gauge field $x$.  This gives
\begin{eqnarray}
    S_{|\mathcal{D}|^{2}}\hspace{-.06in}& = & \hspace{-.06in} \frac{i}{2\pi}\hspace{-.03in}\int_{W} \hspace{-.08in}N_{m} a_{1} \wedge db_{1}  - N_{e}a_{1} \wedge dA_{L} + b_{1} \wedge dx \\
    \hspace{-.06in}& - & \hspace{-.06in}\frac{i}{2\pi}\hspace{-.03in}\int_{W} \hspace{-.08in}N_{m} a_{2} \wedge db_{2}  + N_{e}a_{2} \wedge dx - b_{2} \wedge dA_{R}~.\nonumber 
\end{eqnarray}
Note the change in signs of the second line, which arises from the flip in orientation of the second defect.  Integrating out $x$ enforces the constraint $b_{1}+N_{e}a_{2}=0$.  Hence:
\begin{equation}
    S_{|\mathcal{D}|^{2}}= -\frac{i}{2\pi}\hspace{-.03in}\int_{W}\hspace{-.08in}N_{m}a_{2}\wedge (db_{2}+N_{e}da_{1})+N_{e}a_{1}\wedge dA_{L}+b_{2}\wedge dA_{R}
\end{equation}
Introducing fields $b \equiv -b_2-N_e a_1$, $a\equiv a_{2},$ $c \equiv a_{1}$ we rewrite the above as
\begin{equation}\label{Dsquare}
    S_{|\mathcal{D}|^{2}}=\frac{i }{2\pi} \int _{W} N_{m}a \wedge db + b\wedge dA_{R}+ N_{e}c\wedge (dA_{L}-dA_{R})~.
\end{equation}

To understand \eqref{Dsquare}, let us first consider the special case of $N_m=1$. Then, the equation of motion of the field $a$ trivializes $b$ and $S_{|\mathcal{D}|^{2}}$ simplifies to:
\begin{equation}
   S_{|\mathcal{D}|^2}=\frac{iN_{e}}{2\pi}\int_{W}c \wedge d(A_{L}- A_{R}).
    \label{eq: D square Ne=1}
\end{equation}
Note that if we further set $N_e=1$, the $c$ field acts as a Lagrange multiplier and sets $F_L = F_R$, thus resulting in the trivial operator.  This is as expected: for $N_{e}=N_{m}=1$, $\mathcal{D}_{N_{e}, N_{m}}$ is the $\mathbb{S}$ operator which is invertible.  

\begin{figure}
    \centering
    \includegraphics{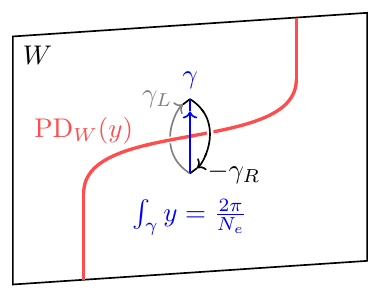}
    \caption{ Illustration of \eqref{jumpformw}. The closed cycle $\gamma_L-\gamma_R$ links with the Poincar\'e dual of $y$ in $W$, and the total holonomy \eqref{jumpformw} takes a fractional value. This implies the presence of an electric one-form symmetry defect (a surface operator) on the Poincar\'e dual of $y$ (red).}
    \label{fig: one-form}
\end{figure}

For a general $N_e$, the action \eqref{eq: D square Ne=1} instead means that $A_{L}$ and $A_{R}$ can differ by a dynamical $\mathbb{Z}_{N_e}$ gauge field $y$ supported on the defect.  We can express this using field $x,y$ and $\lambda$ as
\begin{equation}
    S_{|\mathcal{D}|^2}= \frac{iN_{e}}{2\pi}\int_{W}x \wedge dy + \frac{i}{2\pi} \int_W \lambda \wedge (A_L-A_R-y)~.
    \label{eq: D square Ne=1 mod}
\end{equation}
To understand the meaning of this defect, let us consider how bulk Wilson loops transform.  Pick a segment $\gamma$ in the defect worldvolume $W$, and let $\gamma_{L}$ and $\gamma_{R}$ be slight pushoffs of $\gamma$ to the left and right of $W$ respectively.  The union of $\gamma_{L}$ with $-\gamma_{R}$ is then a bulk one-cycle and we aim to compute the holonomy of the bulk gauge field.  We have the discontinuity:
\begin{equation}\label{jumpformw}
\lim_{\gamma_{i}\rightarrow \gamma}\oint_{\gamma_{L}} A_{L}-\oint_{\gamma_{R}}A_{R}=\oint_{\gamma} y \in \frac{2\pi }{N_{e}}\cdot \mathbb{Z}~.
\end{equation}
For fixed $y$, the discussion below \eqref{1forms} implies that we can interpret \eqref{jumpformw} via the presence of a bulk electric one-form symmetry defect on the two-cycle $S$ in $W$ which is Poincar\'{e} dual to $y$.  See Figure \ref{fig: one-form}.  In the definition of $|\mathcal{D}|^{2}$ we sum over all possible $y$ which is thus equivalent to a sum over all inequivalent insertions of electric one-form symmetry operators on $W$ in the $\mathbb{Z}_{N_{e}}^{(1)}$ group.  

To summarize our result, define $\eta_{e}(S)$ to be the generator of the $\mathbb{Z}_{N_{e}}^{(1)}$ electric one-form symmetry defined on the surface $S$
\begin{equation}
    \eta_{e}(S)=\exp\left(\frac{2\pi i}{N_{e}}Q_{E}(S)\right)=\exp\left(\frac{2\pi }{e^{2}N_{e}}\oint_{S} *F\right)~.
\end{equation}
The fusion derived above is then a sum over all possible such one-form operators:
\begin{equation}
    \mathcal{D}_{N_{e},1}\times  \overline{\mathcal{D}}_{N_{e},1}=\hspace{-.2in}\sum_{S\in H_{2}(W,\mathbb{Z}_{N_e})}\hspace{-.2in}\eta_{e}(S)~.
\end{equation}
In the language of \cite{Gaiotto:2019xmp,Roumpedakis:2022aik}, this is a condensation defect, and we have reproduced the result of \cite{Choi:2021kmx}.

Returning now to the case of general $N_{m}$, \eqref{Dsquare} also includes a $\mathbb{Z}_{N_{m}}$ field $b$ that couples with the magnetic one-form current symmetry current $dA_{R}.$ Therefore summing over $b$ defines a condensation of the magnetic one form symmetry defects.  Introducing $\eta_{m}(S)$ as the generator of the $\mathbb{Z}_{N_{m}}^{(1)}$ subgroup:
\begin{equation}
    \eta_{m}(S)=\exp\left(\frac{2\pi i}{N_{m}}Q_{M}(S)\right)=\exp\left(\frac{i}{N_{m}}\oint_{S} F\right)~,
\end{equation}
we then have the general result:
\begin{equation}\label{fusionfinal}
    \mathcal{D}_{N_{e},N_{m}}\times \overline{\mathcal{D}}_{N_{e},N_{m}}=\hspace{-.2in}\sum_{\substack{  S_{e}\in H_{2}(W,\mathbb{Z}_{N_e})\\ S_{m}\in H_{2}(W,\mathbb{Z}_{N_m})}}\hspace{-.2in}\eta_{e}(S_{e}) \cdot \eta_{m}(S_{m})~.
\end{equation}
The right-hand side of \eqref{fusionfinal} is not the unit operator, thus demonstrating the general non-invertibility of $\mathcal{D}_{N_{e},N_{m}}.$

\subsection{Approximating the Defect}

The defect \eqref{fulldefect} constructed in the previous section is not topological unless the coupling $\frac{e^2}{2\pi}$ is exactly $\frac{N_m}{N_e}$ and in particular rational.  Instead, we can consider the sequence of rational numbers $\frac{N_m^0}{N_e^0}, \frac{N_m^1}{N_e^1}, \cdots$ which approximates $\frac{e^2}{2\pi}$ with increasing precision, resulting in the sequence $\mathcal{D}_j \equiv \mathcal{D}_{N_e^j,N_m^j}$ of non-topological defects.  Here the non-topolgical nature of the defect is quantified by examining the jump in the energy-momentum tensor across the defect as discussed below \eqref{tem}.

Below, we study the relationship between $\mathcal{D}_j$ and $\mathcal{D}_{j+1}$. For concreteness, we assume the sequence is defined by
\begin{equation}
    N_{m}^{j}=2^{j+1}~, ~~~N_e^j=\lfloor\frac{2^{j+2}\pi}{e^2}\rfloor~,
\end{equation}
where above $\lfloor x \rfloor$ denotes the floor of $x$.\footnote{
    Although with this choice $N_m^j$ and $N_e^j$ are not coprime in general, the defect action \eqref{fulldefect} makes sense for such cases. Alternatively, one can also skip those $j$ above with even $N_e^j.$}
Note that if $\frac{e^{2}}{2\pi}$ is rational, the sequence $\mathcal{D}_{j}$ will terminate for sufficiently large finite $j$ in the defect $\mathcal{D}_{N_e^j,N_m^j}$.  Meanwhile, if the coupling $\frac{e^{2}}{2\pi}$, is irrational, the sequence of defects does not terminate and the defects $\mathcal{D}_{j}$ become closer and closer to being topological as $j$ increases.

For each $j$, the defect $\mathcal{D}_j$, supports a 2+1d $\mathbb{Z}_{N_m^j}=\mathbb{Z}_{2^{j+1}}$ gauge theory. To increase the index $j$, we must relate $\mathbb{Z}_{2^{j+1}}$-gauge theory to $\mathbb{Z}_{2^{j}}$-gauge theory.  This can be done using the exact sequence
\begin{equation}
    1 \to \mathbb{Z}_{2} \to \mathbb{Z}_{2^{j+1}} \to \mathbb{Z}_{2^j}\to 1~.
\end{equation}
Explicitly, instead of the standard action for $\mathbb{Z}_{2^{j+1}}$ gauge theory:
\begin{equation}
    \frac{2^{j+1}i}{2\pi}\int a^{(j)}d b^{(j)}~,
    \label{eq: Z2j}
\end{equation}
we can use the following action
\begin{equation}
    \frac{i}{2\pi}\int_{W} 2^{j}a^{(j-1)}\wedge d {b}^{(j-1)}+ 2\tilde{a}^{(j)}\wedge d \tilde{b}^{(j)} -  a^{(j-1)}\wedge d \tilde{b}^{(j)}~.
    \label{eq: Z2j decompose}
\end{equation}
To go from \eqref{eq: Z2j decompose} to \eqref{eq: Z2j}, consider the following change of variables in $GL(4,\mathbb{Z})$:
\begin{align}
    \tilde a^{(j)} = a^{(j)}~, \quad & b^{(j-1)} =  b^{(j)}~, \\
    2 \tilde a^{(j)} - a^{(j-1)} =  c~, \quad  &2^j b^{(j-1)} - \tilde b^{(j)}= e~. \nonumber
\end{align}
This results in the action $\eqref{eq: Z2j}$ together with a coupling $\frac{i}{2\pi} \int c \wedge d e$ trivializing $c$ and $e$.  We can also incorporate the coupling of $a^{(j)}$ and $b^{(j)}$ to the bulk electromagnetic field in \eqref{fulldefect} by adding
\begin{equation}
     \frac{i}{2\pi} \int b^{(j-1)}\wedge dA_R-\frac{i}{2\pi}\int (N^{j,j-1}_e\tilde{a}^{(j)}+N^{j-1}_e a^{(j-1)})\wedge dA_L~,
    \label{eq: decomposed bulk coupling}
\end{equation}
to \eqref{eq: Z2j decompose}, where $N^{j,j-1}_e = N^j_e - 2 N^{j-1}_e$ is the last bit of $N_e^j$.

Thus, we have found that for this particular choice of $(N^j_e,N^j_m)$ the defects $\mathcal{D}_j$ and $\mathcal{D}_{j+1}$ are related as
\begin{equation}
    \mathcal{D}_{j+1} \cong \mathcal{D}_{j}\otimes(\text{$\mathbb{Z}_2$ gauge theory})~,
    \label{eq: D decompose}
\end{equation}
where the two factors in the right hand side have the coupling of the last term of \eqref{eq: Z2j decompose}, and the $\mathbb{Z}_2$ gauge theory also couples to the bulk via the first term of \eqref{eq: decomposed bulk coupling}.

One can iterate the decomposition \eqref{eq: D decompose} to rewrite \eqref{fulldefect} as  $j+1$ coupled copies of $\mathbb{Z}_2$ gauge theory. The total action becomes:
\begin{eqnarray}
S_{\mathcal{D}_{j}}& = &    \frac{i}{2\pi}\int_{W}\left(~\sum_{k=0}^j 2 \tilde{a}^{(k)}\wedge d\tilde{b}^{(k)}
    -\sum_{k=1}^j\tilde{a}^{(k-1)}\wedge d\tilde{b}^{(k)}\right. \nonumber\\
    & - & \left.\sum_{k=0}^j N_e^{k,k-1}\tilde{a}^{(k)}\wedge dA_L
    +\tilde{b}^{(0)}\wedge dA_R\right)~,
\end{eqnarray}
where $N^{0,-1}_e$ is defined as $N^0_e$.  As discussed above, for rational coupling the above terminates in the defect $\mathcal{D}_{N_{e},N_{j}}.$  Meanwhile, for $\frac{e^{2}}{2\pi}$ irrational, the $j\to \infty$ limit formally defines a topological defect which supports infinitely many species of anyons.

\section{Non-Invertible Symmetry Action}

The interface definition \eqref{fulldefect} of the defect $\mathcal{D}_{N_{e},N_{m}}$ allows us to explore the action of the duality defect on local and extended operators.  

\subsection{Action on Operators}

To begin,  the continuity equation \eqref{continuity} shows that across the defect $iF_{L}$ exchanges with $*F_{R}$.  Since the field strengths generate the gauge invariant local operators, this fixes the action of $\mathcal{D}_{N_{e},N_{m}}$ on all local operators.  For instance, the gauge field kinetic term is odd under the action of $\mathcal{D}_{N_{e},N_{m}}:$
\begin{equation}
  \mathcal{D}_{N_{e},N_{m}}:  F_{L}\wedge *F_{L}\rightarrow -F_{R}\wedge *F_{R}~.
\end{equation}
Thus as expected, the symmetry $\mathcal{D}_{N_{e},N_{m}}$ implies that the coefficient of the kinetic term, i.e.\ the coupling constant cannot be changed infinitesimally.

We can also easily determine the action of $\mathcal{D}_{N_{e},N_{m}}$ on extended operators.  Indeed, the operators $*F$ and $F$ are also the currents for the electric and magnetic one-form global symmetries.  Thus we have:
\begin{equation}
    \mathcal{D}_{N_{e},N_{m}}(Q_{E}^{R})=\frac{N_{e}}{N_{m}}Q_{M}^{L}~, ~~~\mathcal{D}_{N_{e},N_{m}}(Q_{M}^{R})=-\frac{N_{m}}{N_{e}}Q_{E}^{L}~,
\end{equation}
which generalizes the action of $\mathbb{S}$ duality in \eqref{smap}.

Relatedly, we can also consider operators that define heavy probe particles carrying electric and magnetic charges $(q_{e},q_{m}).$  If both the charges $q_{i}$ are integral then these operators are described by line operators, i.e.\ Wilson/'t Hooft lines or more general dyons.  If the the charges are fractional then these particles have visible Dirac strings and may be properly understood as worldines that are tied to the boundary of open topological surfaces.   Composing the one-form symmetry gauging with the action \eqref{smap} of electric magnatic duality on charges gives
\begin{equation}
    \mathcal{D}_{N_{e},N_{m}}\left[(q_{e},q_{m})\right]=\left(\frac{q_{m}N_{e}}{N_{m}}, -\frac{N_{m}q_{e}}{N_{e}}\right)~.
\end{equation}
Note that this transformation in general maps integral charges, corresponding to lines, to fractional charges corresponding to open surfaces.  Such exchanges between ``genuine" line operators and open boundaries is a universal feature of non-invertible symmetries.

\subsection{Action on Boundary Conditions}\label{bcsec}

We can also consider the action of the defect on boundary conditions.  We can obtain the action on these boundary conditions using \eqref{fulldefect}.  By sandwiching the duality defect action with the boundary action we obtain a new boundary with $A_{L}$ interpreted as a new boundary gauge field and $A_{R}$ interpreted as the remaining bulk field.  We let $\mathcal{B}$ denote a given boundary condition, and indicate the resulting boundary condition as $\mathcal{B}\cdot \mathcal{D}_{N_{e},N_{m}}.$  (See Figure \ref{fig: one-form}, with $\mathbb{S}$ replaced by $\mathcal{D}_{N_e,N_m}.$ ) The analysis below generalizes that of \cite{Kapustin:2009av} which studied the analogous problem for $\mathbb{S}.$

\emph{Action on Perfect Electric Conductor}

Consider first the case where $\mathcal{B}$ is a perfect electric conductor \eqref{PEC}.  Acting with the symmetry defect on a perfect electric conductor yields the boundary action:
\begin{equation}\label{fulldefectact}
 S_{b}   =\frac{i}{2\pi}\int_{\partial M} N_{m} a \wedge db+b \wedge dA-N_{e}a \wedge da'+c \wedge da'~.
\end{equation}
Simplifying using the equation of motion of $c$ yields the result:
\begin{equation}\label{PECdual}
 S_{b} =  \frac{iN_{m}}{2\pi}\int_{\partial M} a \wedge db +\frac{i}{2\pi}\int_{\partial M} b \wedge dA~.
\end{equation}
The boundary fields $a$ and $b$ describe a $\mathbb{Z}_{N_{m}}$ topological gauge theory which couples to the bulk electromagnetic field through its one-form symmetry.  The equations of motion yield:
\begin{eqnarray}
   N_{m}\frac{db}{2\pi}& = & 0~. \nonumber \\
   \frac{1}{ie^{2}}*F|_{\partial M}+\frac{db}{2\pi} &=&0~. \\
\frac{F}{2\pi}|_{\partial M}+N_{m}\frac{da}{2\pi}&=&0~. \nonumber
\end{eqnarray}   
For $N_{m}=1$, the first two equations set $b$ to be trivial and $*F|_{\partial M}$ to vanish yielding a perfect magnetic conductor consistent with the discussion around \eqref{PECs}.  More generally, for $N_m>1,$  $b$ is a $\mathbb{Z}_{N_{m}}$ gauge field and the bulk electric one-form symmetry is restricted to be the Bockstein $\beta(b)\in H^{2}(\partial M, \mathbb{Z})$, which is necessarily a torsion class. (See \eqref{bockdef}.)

\emph{Action on Perfect Magnetic Conductor}

Analogously, if the boundary condition $\mathcal{B}$ is the perfect magnetic conductor \eqref{pmccond}, the defect action yields a boundary condition $  \mathcal{B}\cdot \mathcal{D}_{N_{e},N_{m}}$ with a boundary action:
\begin{equation}
    S_{b}=\frac{iN_{m}}{2\pi}\int a \wedge db+\frac{i}{2\pi}\int b \wedge dA-\frac{iN_{e}}{2\pi}\int a \wedge dc~.
\end{equation}
By integrating out $b$ and $c$ we deduce the boundary equations of motion:
\begin{align}
    N_{e}\frac{da}{2\pi}  &=  0~, \nonumber \\
    N_{m}\frac{da}{2\pi}+\frac{F}{2\pi}|_{\partial M}  &=  0~,\\
     \frac{1}{ie^{2}}*F|_{\partial M}+\frac{db}{2\pi} &= 0~. \nonumber
\end{align}
In the special case $N_{e}=1,$ the first equation trivializes $a$ and the second equation yields a perfect electric conductor consistent with \eqref{PECs}.  Meanwhile for general $N_{e},$ $a$ is restricted to be a $\mathbb{Z}_{N_{e}}$ gauge field propagating on the boundary. The boundary flux $F|_{\partial M}$ need not vanish but instead is restricted to be the torsion class $\beta(a).$


\subsection{Toroidal Cavity Hilbert Spaces}

To illustrate the concepts above let us examine the duality symmetry acting on Hilbert spaces.  We again take as an example the toroidal cavity Hilbert space described in Section \ref{sechilb}.  At the boundary we place a given boundary condition $\mathcal{B}$.  

Along the non-contractible one-cycle in the solid torus we place a condensate of one-form symmetry defects extending along time. Specifically, we insert the operator:
\begin{equation}
  \mathcal{C}(S^{1}\times \mathbb{R}) \equiv \sum_{i=1}^{N_{e}}\sum_{j=1}^{N_{m}}\eta_{e}(S^{1}\times \mathbb{R})^{i} \times \eta_{m}(S^{1}\times \mathbb{R})^{j}~.
\end{equation}
Since this is an insertion along time, we may view it as a modification of the Hilbert space to the defect Hilbert space $\mathcal{H}_{\mathcal{C}, \mathcal{B}},$ where the subscript $\mathcal{B}$ indicates that these states also depend on the boundary condition.  In this setup the meaning of the sum above is that the Hilbert space $\mathcal{H}_{\mathcal{C}, \mathcal{B}}$ is a direct sum of sectors, each of which arises from an insertion of a single one-form symmetry defect along time. 

\begin{figure}
    \centering
    \begin{minipage}[]{.4\linewidth}
    \centering
    \includegraphics[width=\linewidth]{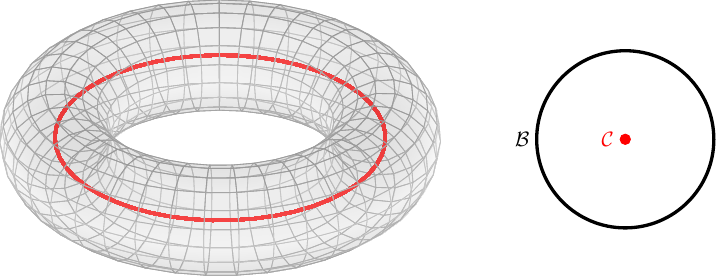}
    \subcaption{}    
    \end{minipage}
    \hspace{.08\linewidth}
    \begin{minipage}[]{.4\linewidth}
    \centering
    \includegraphics[width=\linewidth]{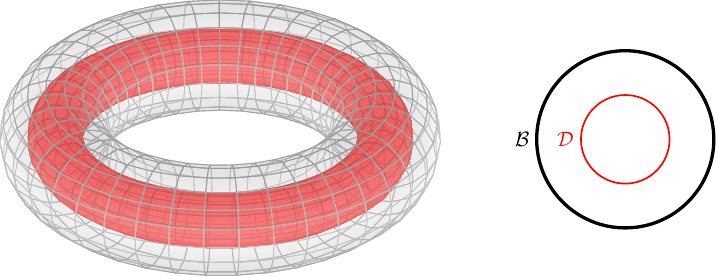}
    \subcaption{}    
    \end{minipage}\\
    \begin{minipage}[]{.4\linewidth}
    \centering
    \includegraphics[width=\linewidth]{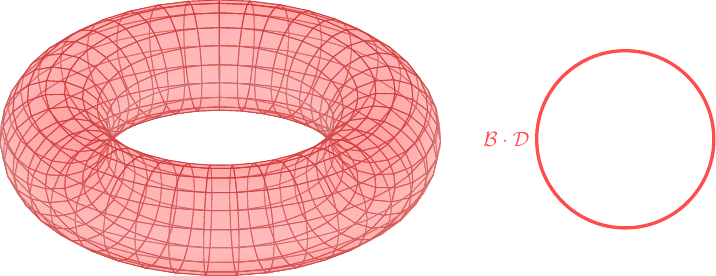}
    \subcaption{}    
    \end{minipage}
    \caption{The topological move that proves \eqref{eq: t cavity equivalence}. Both the toroidal cavity and a cross-section are illustrated.
    (a): The configuration defining $\mathcal{H}_{\mathcal{C},\mathcal{B}}$. The red ring is the condensate $\mathcal{C}$, and the toroidal cavity has the boundary condition $\mathcal{B}$.
    (b): The condensate $\mathcal{C}$ is replaced by the toroidal duality defect $\mathcal{D}=\mathcal{D}_{N_e,N_m}$ using the fusion rules for $\mathcal{D}\times \overline{\mathcal{D}}$.
    (c): The duality defect is topological and can be pushed onto the boundary, changing the boundary condition to $ \mathcal{B} \cdot \mathcal{D}$.}
    \label{fig:enter-label}
\end{figure}

However, there is an alternative way to understand the operator $\mathcal{C}$ that will prove fruitful below.  Specifically, we claim that we can replace the insertion of $\mathcal{C}$ with a duality defect along a small torus at the center of the geometry:
\begin{equation}\label{nucleate}
    \mathcal{D}_{N_{e},N_{m} }(T^{2}\times \mathbb{R})= \mathcal{C}(S^{1}\times \mathbb{R})~.
\end{equation}
To argue for \eqref{nucleate}, we use the fusion rule \eqref{fusionfinal} to contract the small bubble of duality defect to a condensation operator.  To see this equivalence we can deform the contractible cycle on the duality defect until it appears to be locally two oppositely oriented segments.  Colliding these segments using the fusion rule we obtain a codimension one condensation defect of one form-symmetry defects on $S^{1}\times \mathbb{R} \times I$, where $I$ is an interval arising from fusing the two sides of the circle.  At the ends of this interval the two-form gauge fields defining the condensation vanish and thus the sum is over one-form symmetry operators labelled by two cocycles in the relative cohomology group:
\begin{equation}
   H^{2}(S^{1}\times \mathbb{R} \times I, \partial(S^{1}\times \mathbb{R} \times I), \mathbb{Z}_{N_{i}})\cong H_{2}(S^{1}\times \mathbb{R}, \mathbb{Z}_{N_{i}})~,
\end{equation}
where in the final step we used Lefshetz duality.  Applying this to each of the $\mathbb{Z}_{N_{i}}$ factors separately, then shows that a small torus shaped duality defect is equivalent to an insertion of $\mathcal{C}$.  See Figure \ref{fig:enter-label}.

\newpage

\onecolumngrid

\twocolumngrid

We can use the equivalence \eqref{nucleate} to study the Hilbert space $\mathcal{H}_{\mathcal{C}, \mathcal{B}}$ in two different ways:
\begin{itemize}
    \item Directly evaluate the defect Hilbert space $\mathcal{H}_{\mathcal{C}, \mathcal{B}}$ by appropriately quantizing in the presence of the insertion $\mathcal{C}$ and boundary condition $\mathcal{B}$
      \item Expand the duality defect $\mathcal{D}_{N_{e},N_{m} }(T^{2}\times \mathbb{R})$ until it collides with the boundary $\mathcal{B}$.  Using the methods discussed in Section \ref{bcsec} this results in a modified boundary condition $ \mathcal{B}\cdot \mathcal{D}_{N_{e},N_{m} }$, but with no additional insertion of one-form symmetry defects.  
\end{itemize}
In summary, the argument above implies that at rational coupling, we have the equivalence of Hilbert spaces and energy levels:
\begin{equation}
    \mathcal{H}_{\mathcal{C}, \mathcal{B}} \cong \mathcal{H}_{1,   \mathcal{B}\cdot \mathcal{D}_{N_{e},N_{m} }}~,
    \label{eq: t cavity equivalence}
\end{equation}
where in the right-hand side the subscript $1$ indicates that this is the untwisted Hilbert space.  

Below we illustrate this correspondence in a sector of the Hilbert space for the case when $\mathcal{B}$ is a perfect electric conductor.

\subsubsection{Twisted Sector Hilbert Spaces}

Let us first consider the case with the insertion of $\mathcal{C}$ and perfect electric conductor boundary conditions.  Since the one-form symmetry defects defining $\mathcal{C}$ are topological, we can deform their location without altering the problem.  In particular, consider the magnetic insertions which are exponentials of $\int_{S^{1}_{\phi}\times \mathbb{R}} F$.  By deforming the location of the $\phi$ circle from $r=0$, the center of the solid torus, to $r=R,$ the boundary, we can use the boundary conditions to trivialize these insertions.  Therefore we have the simplification:
\begin{equation}\label{dubdirectsum}
    \mathcal{H}_{\mathcal{C}, \mathcal{B}} \cong \oplus_{i=1}^{N_{m}}\oplus_{p=1}^{N_{e}} \mathcal{H}_{\eta_{e}(S^{1}_{\phi}\times \mathbb{R})^{p},\mathcal{B}}~.
\end{equation}

We again consider the class of modes where the only non-vanishing components of the gauge field are $A_{\theta}$. (For simplicity of notation we also denote this subspace by $\mathcal{H}$.)  As in \eqref{d1sol} the solution to the equation of motion obeying the Dirichlet boundary condition is:
\begin{equation}
    A_{\theta}=\beta (r^{2}-R^{2})~.
\end{equation}
Now the circle parameterized by $\theta$ links the one-form symmetry defect $\eta_{e}(S^{1}_{\phi}\times \mathbb{R})^{p}$.  Therefore the solution must have a prescribed non-trivial holonomy as $r\rightarrow 0$:
\begin{equation} 
\lim_{r\rightarrow0}\exp\left(i\oint A \right)=\exp\left(\frac{2\pi i p}{N_{e}}\right)~.
\end{equation}
This quantizes the parameter $\beta$ yielding:
\begin{equation}\label{athtwisted}
    A_{\theta}=\left(n-\frac{p}{N_{e}}\right)\left(\frac{r^{2}}{R^{2}}-1\right)~,~~~n\in \mathbb{Z}~.
\end{equation}
Thus, the flux through the disc is now fractional, with the fractional part specified by which twisted sector the state is in.  We again use the action \eqref{maxact}, to evaluate the energy of these states yielding:
\begin{equation}
    E_{n,p}=\frac{4\pi^{2}L}{e^{2}R^{2}}\left(n-\frac{p}{N_{e}}\right)^{2}=\frac{2\pi L N_{e}}{R^{2}N_{m}}\left(n-\frac{p}{N_{e}}\right)^{2}~.
\end{equation}

Finally, we can determine the full $\mathcal{H}_{\mathcal{C},\mathcal{B}}$ Hilbert space by summing over $p$.  This simply means that the flux through the disc defined via \eqref{athtwisted} can be any fraction with denominator $N_{e}$.  Hence we have:
\begin{equation}\label{sectorsum}
    \mathcal{H}_{\mathcal{C}, \mathcal{B}} \cong \oplus_{k=0}^{\infty} \mathcal{H}_{k}~,
\end{equation}
where the subspace $\mathcal{H}_{k}$ for $k\in \mathbb{Z}$ is those states with flux $k/N_{e}$ through the disc. Each such sector has degeneracy $N_{m}$ (arising from the first direct sum in \eqref{dubdirectsum}) with energy $E_{k}$ given by:
\begin{equation}\label{estwisted}
    E_{k}=\left(\frac{2\pi L }{R^{2}N_{e}N_{m}}\right)k^{2}~.
\end{equation}

\subsubsection{Modified Boundary Conditions}

Alternatively, we can modify the boundary condition by the action of the defect $\mathcal{D}_{N_{e},N_{m}}.$  This results in Maxwell theory at the same coupling $\frac{e^{2}}{2\pi}=\frac{N_{m}}{N_{e}}$ but now coupled to a boundary $\mathbb{Z}_{N_{m}}$ topological gauge theory through the action \eqref{fulldefectact}.  

Let us first discuss the topological sector.  This theory is quantized on the boundary torus and as such has ground states in one-to-one correspondence with its distinct line operators.  In this case there are $N_{m}^{2}$ such line operators, and the associated states may be labelled by specifying definite holonomy of the gauge fields $a$ and $b$ in \eqref{fulldefectact} around a fixed cycle.  For convenience, we choose this to be the $\theta$-cycle.  We indicate these states as $|p_{a},p_{b}\rangle$ with $p_{i}=0, 1, \cdots N_{m}-1$:
\begin{eqnarray}
     \left(\oint_{\phi=\phi_{o}} \hspace{-.1in}a~\right)|p_{a},p_{b}\rangle & = & \frac{2\pi p_{a}}{N_{e}}|p_{a},p_{b}\rangle~,\\
     \left(\oint_{\phi=\phi_{o}} \hspace{-.1in}b~\right)|p_{a},p_{b}\rangle  & = & \frac{2\pi p_{b}}{N_{e}}|p_{a},p_{b}\rangle~. \nonumber
\end{eqnarray}

In each of these states we can examine the coupling to the bulk gauge field.  While the $a$ field does not couple with the bulk and thus the eigenvalue $p_a$ does not affect the energy, the $b$ field couples to the boundary magnetic symmetry in \eqref{PECdual}.  Considering as in Section \ref{sechilb} the modes with only $A_{\phi}$ non-trivial but time dependent we arrive at the action
\begin{equation}
    S=\frac{\pi^{2} R^{2}}{Le^{2}}\int \dot{A}_{\phi}^{2} ~ dt+\frac{2\pi i p_{b}}{N_{m}}\int \dot{A}_{\phi} ~ dt~.
\end{equation}
This generalizes \eqref{n1sol} to the action for a quantum particle on a circle now in the presence of a magnetic field ($\theta$-term) specified by $p_{b}.$

The spectrum is again given by quantizing the canonical momentum
\begin{equation}
P=\frac{2\pi^{2}R^{2}}{Le^{2}}\dot{A}_{\phi}+\frac{2\pi p_{b}}{N_{m}} =2\pi n~,~~~n\in \mathbb{Z}~,
\end{equation}
resulting in energy in states $|p_{a},p_{b},n\rangle$ with energies:
\begin{equation}
    E_{p_{a}, p_{b},n}=\left(\frac{e^{2}L}{R^{2}}\right)\left(n-\frac{p_{b}}{N_{m}}\right)^{2}~.
\end{equation}
Since the Hilbert space includes a direct sum over $p_{b}$, we see that all possible fractions with denominator $N_{m}$ are achieved above.  Thus the Hilbert space again decomposes into sectors as in \eqref{sectorsum} where each sector has degeneracy $N_{m}$ (arising from the sum over $p_{a}$) and the energies are given by:
\begin{equation}
    E_{k}=\left(\frac{e^{2}L}{R^{2}N_{m}^{2}}\right)k^{2}=\left(\frac{2\pi L}{R^{2}N_{e}N_{m}}\right)k^{2}~.
\end{equation}
Happily, this agrees exactly with \eqref{estwisted}.



\let\oldaddcontentsline\addcontentsline
\renewcommand{\addcontentsline}[3]{}
\section*{Acknowledgements}
\let\addcontentsline\oldaddcontentsline

We thank S. Koren for discussions.
CC is supported by the US Department of Energy DE-SC0009924 and the Simons Collaboration on Global Categorical Symmetries.
KO is supported by JSPS KAKENHI Grant-in-Aid No.22K13969 and the Simons Collaboration on Global Categorical Symmetries.


\let\oldaddcontentsline\addcontentsline
\renewcommand{\addcontentsline}[3]{}
\bibliography{duality}
\let\addcontentsline\oldaddcontentsline

\end{document}